\pdfoutput=1
\documentclass[conference]{IEEEtran}
\IEEEoverridecommandlockouts

\usepackage{cite}
\usepackage{amsmath,amssymb}
\usepackage{graphicx}
\usepackage{booktabs}
\usepackage{xcolor}
\usepackage{pgfplots}
\usepackage{url}
\usepackage{microtype}
\pgfplotsset{compat=1.18}
\usetikzlibrary{arrows.meta,positioning,fit,backgrounds}

\usepackage[hidelinks]{hyperref}

\definecolor{pblue}{HTML}{2B5C8F}
\definecolor{pteal}{HTML}{2E8B57}
\definecolor{pamber}{HTML}{D97706}
\definecolor{pgray}{HTML}{F1F5F9}

\newcommand{\CHIP}{a complex multi-architecture, multi-core SoC}
\newcommand{\CORELIST}{multiple heterogeneous 64-bit and 32-bit application, real-time, and embedded DSP cores spanning Arm, RISC-V, and VLIW/SIMD DSP instruction-set architectures}
\newcommand{\SIMULATOR}{a commercial event-driven SystemVerilog RTL simulator}
\newcommand{\TUBE}{log tube}
\newcommand{\DICTFILENAME}{\texttt{string\_\allowbreak dict.txt}}
\newcommand{\TOKDEFINE}{\texttt{-DTOKENIZED\_LOG}}
\newcommand{\SUITEA}{Subsystem Suite~A}
\newcommand{\SUITEB}{Subsystem Suite~B}
\newcommand{\CASEA}{Workload~A1: Address Fault Handling (433p)}
\newcommand{\CASEB}{Workload~A2: Coherency Stress (106p)}
\newcommand{\CASEC}{Workload~A3: Concurrent Fabric Traffic (81p)}
\newcommand{\CASED}{Workload~A4: Doorbell Interrupt Path (85p)}
\newcommand{\CASEE}{Workload~A5: Low-Power UPF Retention (96p)}
\newcommand{\CASEF}{Workload~A6: Firmware Boot + Full Waveform (49p)}
\newcommand{\CASEG}{Workload~B1: Low-Power Boot Access (312p)}
\newcommand{\CASEH}{Workload~B2: Peripheral Serial Stress (148p)}
\newcommand{\CASEI}{Workload~B3: Peripheral UART Mux (84p)}

\newcommand{\Clegacy}{C_{\mathrm{legacy}}}

\newcommand{\Ctoken}{C_{\mathrm{token}}}

\newcommand{\Cfmt}{C_{\mathrm{fmt}}}

\newcommand{\Cwr}{C_{\mathrm{wr}}}

\newcommand{\Cbar}{C_{\mathrm{bar}}}

\newcommand{\Cpush}{C_{\mathrm{push}}}

\newcommand{\msglen}{L}

\newcommand{\nargs}{n}

\newcommand{\nslots}{S}

\newcommand{\density}{\rho}

\newcommand{\speedup}{\sigma}

\begin{document}

\title{Stop Paying by the Letter: Accelerating Full-Chip RTL Simulation with Host-Side Tokenized Logging}

\author{\IEEEauthorblockN{Daniel Keller and Rohit Kumar}
\IEEEauthorblockA{Google\\
\{kellerdaniel, rohitpipania\}@google.com}}

\maketitle

\begin{abstract}
Firmware diagnostic logging (\texttt{printf}) is an essential debug instrument
in full-chip system-on-chip (SoC) design verification, yet it imposes a heavy
tax on cycle-accurate RTL simulation: formatting a $30$-character, $1$-argument
message on a simulated CPU and serializing it through a memory-mapped aperture
costs $3.04$ to $15.23\,\mu\text{s}$ per call (up to $19.60\,\mu\text{s}$ for a
$34$-character, $4$-argument call; $15.27\,\mu\text{s}$ unweighted / $15.08\,\mu\text{s}$
print-weighted critical-path mean across \SUITEA{}). Unlike embedded tokenizers that target flash footprint and serial
wire bytes with offline decoding, RTL simulation is governed by a two-regime
cost function---cross-fabric SoC paths are dominated by MMIO bus stalls
($84.7\%$ of call latency) whereas local subsystem paths are dominated by
on-core \texttt{vsprintf} formatting ($61.7\%$)---and requires live,
zero-simulation-time string reconstruction to drive UVM scoreboards.

We present a host-side in-simulation tokenized logging architecture for
heterogeneous multi-core SoCs that eliminates both bottlenecks with zero
firmware call-site edits. Using link-time ELF format-string virtual addresses as
tokens over a fixed $32$-bit word-aligned slot protocol (identical across Arm,
RISC-V, and DSP cores), on-core logging collapses to a single-pass specifier
scan and $(1+\nslots)$ MMIO word writes. This removes all on-core radix
conversion ($\Cfmt \to 0$), $85.0\%\text{--}94.5\%$ of deployed bus transactions
($6.7\times\text{--}18.3\times$), $84.6\%\text{--}93.3\%$ of MMIO bus-stall
cycles, and $63.5\%\text{--}90.9\%$ of total on-core logging cycles
($2.74\times\text{--}11.03\times$; $69.6\%\text{--}70.4\%$ of retired
instructions), while a SystemVerilog monitor reconstructs bit-exact strings in
zero simulated cycles ($\Delta t_{\mathrm{sim}} = 0$). Across $858$ matched
full-chip test pairs ($1{,}716$ paired runs across $1{,}972$ total runs) spanning two production suites
(\SUITEA{}, $N=169$; \SUITEB{}, $N=689$), we achieve $100\%$ output parity with
zero simulator memory overhead, saving $1{,}314.85$ simulator CPU hours
($-26.18\%$) per regression pass, reducing median single-test CPU time by
$41.51\%$ on \SUITEA{} and $21.33\%$ on \SUITEB{}, boosting mean effective
simulator throughput by $+24.02\%$ on \SUITEA{} and $+19.60\%$ on \SUITEB{} per
CPU second ($+8.51\%$ and $+23.25\%$ per wall-clock second), and shrinking
\SUITEA{} instruction traces by $42.84\%$ and waveform dumps by $28.82\%$
(medians).
\end{abstract}


\section{Introduction}
\label{sec:intro}

Consider a single line of firmware diagnostic output emitted thousands of times
during full-chip system-on-chip (SoC) bring-up and verification:

\begin{center}
\texttt{[SYS\_INIT] Subsys channel ready (status=0x00, retry=03)}
\end{center}

On physical silicon, emitting this message takes microseconds and is negligible.
Inside a cycle-accurate RTL simulation of that same silicon, however, it is
computationally expensive. Under a standard bare-metal C library, the simulated
processor executes hundreds of instructions of integer-to-ASCII conversion
(\texttt{vsprintf}) through its fetched, decoded, and retired pipeline stages,
followed by dozens of memory-mapped I/O (MMIO) bus transactions across the
interconnect fabric---and, on one of the three subsystems we study, a full
data-synchronization barrier after every character. Across the production SoC
topologies evaluated in Section~\ref{sec:costmodel} and Section~\ref{sec:eval},
a single $30$-character, $1$-argument print call consumes $3.04\,\mu\text{s}$ on
a local subsystem tube to $15.23\,\mu\text{s}$ on a cross-fabric SoC tube
(Table~\ref{tbl:trace_decomp}; up to $19.60\,\mu\text{s}$ for a $34$-character,
$4$-argument call), and saves a critical-path mean of
$\Delta C_{\mathrm{sim}} = 15.27\,\mu\text{s}$ unweighted ($15.08\,\mu\text{s}$
print-weighted) of simulated time per print across full-chip multi-core tests on
\SUITEA{} (\textbf{$\approx 15{,}080\text{--}15{,}270$ cycles at $1\,\text{GHz}$}, vs.\
$3.64\text{--}5.41\,\mu\text{s}$ on peripheral-dominated \SUITEB{} workloads). Because full-chip event-driven RTL
simulation executes millions of times slower than silicon, simulating those
logging cycles burns \textbf{$37.8$ to $441.5$ host server CPU seconds ($7.36$
CPU minutes in power-aware UPF simulation) per \texttt{printf} call}.

\subsection*{The Simulation Cost Asymmetry}

Deferring string formatting by transmitting a compact \emph{token} is well
established in embedded software~\cite{pwtokenizer,nanolog2018,zephyrdict,trice,defmt},
where the scarce resources are \emph{flash memory} and \emph{serial wire bytes}
and logs are decoded \emph{offline}. Cycle-accurate RTL simulation obeys two
different constraints:
\begin{enumerate}
  \item \textbf{Transactions and formatting dominate in two topology-dependent
    regimes, not byte count:} On cross-fabric SoC paths, interconnect handshakes
    account for $84.7\%$ of call latency ($430.0\,\text{ns/char}$ once the
    $3$-entry store buffer fills), whereas on local subsystem paths
    ($27.0\,\text{ns}$ per barriered write), software \texttt{vsprintf}
    formatting accounts for $61.7\%$ (Table~\ref{tbl:trace_decomp}). Variable-length
    byte packing adds CPU instructions without saving bus transactions; a fixed
    $32$-bit slot protocol removes on-core radix conversion ($\Cfmt \to 0$),
    cuts interconnect write transactions by $6.7\times\text{--}18.3\times$
    ($85.0\%\text{--}94.5\%$), and reduces MMIO bus-stall cycles by
    $84.6\%\text{--}93.3\%$.
  \item \textbf{Reconstruction must occur in-simulation in zero simulated time:}
    Firmware logs actively drive live SystemVerilog/UVM scoreboards,
    severity-based error counters, and real-time pass/fail adjudication, so
    strings must materialize dynamically at the exact simulation timestamp they
    are emitted ($\Delta t_{\mathrm{sim}} = 0$).
\end{enumerate}

\subsection*{Host-Side Tokenized Logging for RTL Verification}

We present a host-side in-simulation tokenized logging architecture for
heterogeneous multi-core SoCs. Using the link-time read-only memory
(\texttt{.rodata}) virtual address of each format string as its token ID
intercepts standard C \texttt{printf} calls transparently without modifying a
single call site. Because the protocol relies solely on ELF section tables, C99
variadic promotion, and 32-bit MMIO stores, it operates identically across Arm
(\texttt{AArch64}/\texttt{AArch32}), RISC-V (\texttt{RV32}/\texttt{RV64}), and
embedded DSP cores. On the simulated processor, radix conversion is eliminated
($\Cfmt \to 0$), leaving only a single-pass specifier scan ($\Cpush$) and
$1+\nslots$ word writes—cutting total on-core logging cycles by
$63.5\%\text{--}90.9\%$ ($2.74\times\text{--}11.03\times$) and retired
instructions by $69.6\%\text{--}70.4\%$—while a SystemVerilog monitor resolves
the token against a build-time sidecar dictionary and emits bit-exact UVM log
messages in zero simulated clock cycles ($\Delta t_{\mathrm{sim}} = 0$).

\subsection*{Summary of Contributions}

\begin{itemize}
  \item \textbf{Two-Regime Cost Model \& Effective Throughput Analysis:} We
    derive a serial-fraction cost model (Section~\ref{sec:costmodel}) grounded
    in instruction-retirement trace decomposition across three SoC logging
    topologies ($\Delta C = 1.93\text{--}13.85\,\mu\text{s}$ at $\msglen=30,
    \nslots=1$) and show that excising high-toggle logging windows increases
    mean effective simulator throughput by $+24.02\%$ on \SUITEA{} and
    $+19.60\%$ on \SUITEB{} per CPU second ($+8.51\%$ on \SUITEA{} [damped by a
    fixed $\approx 1.25\,\text{h}$ flush tail] and $+23.25\%$ on \SUITEB{} per
    wall-clock second).
  \item \textbf{Zero-Simulation-Time In-Testbench Reconstruction:} We design a
    modal SystemVerilog/UVM token decoder that reconstructs formatted strings in
    zero simulated cycles ($\Delta t_{\mathrm{sim}} = 0$) with $0.00\%$
    simulator memory overhead (Section~\ref{sec:design}).
  \item \textbf{ISA-Independent Zero-Call-Site-Edit Integration:} Link-time ELF
    virtual-address tokens and a normalized 32-bit wire slot contract enable
    drop-in deployment across Arm, RISC-V, and DSP subsystems via a single
    header alias (Section~\ref{sec:impl}).
  \item \textbf{Production Full-Chip Evaluation ($N=858$ Pairs):} Across $858$
    matched full-chip test pairs ($1{,}716$ paired runs across $1{,}972$ total runs) spanning \SUITEA{}
    ($N=169$) and \SUITEB{} ($N=689$) on \CHIP{} (Section~\ref{sec:eval}), our
    architecture achieves $100\%$ bit-exact output parity, saves \textbf{$1{,}314.85$
    simulator CPU hours ($-26.18\%$)} and \textbf{$1{,}670.75$ wall-clock hours
    ($-27.09\%$)} per regression pass, reduces median single-test CPU time by
    \textbf{$41.51\%$ ($1.706\times$ GMean, up to $5.94\times$)} on \SUITEA{}
    and \textbf{$21.33\%$ ($1.266\times$ GMean)} on \SUITEB{}, and shrinks
    compressed waveform dumps by \textbf{$28.82\%$ median ($2.98\text{--}5.94$\,GB
    per deep debug trace)}.
  \item \textbf{Exposing Latent Firmware Timing Races:} Removing microseconds of
    artificial per-print serialization delay surfaces latent multi-core firmware
    synchronization races—such as an unsynchronized $6$-processor CSR
    read-modify-write race with a $3\%$ ($261\,\text{ns}$) margin
    (Section~\ref{sec:latent})—that character-serialized logging had previously
    masked.
\end{itemize}

\section{Background and Motivation}
\label{sec:background}

\subsection{Firmware Logging in Cycle-Accurate RTL Simulation}

Full-chip SoC verification executes production bare-metal bootcode and firmware
directly on simulated processor RTL. Because physical UART peripherals require
millions of simulated clock cycles per character at standard baud rates,
verification environments provide a memory-mapped simulation logging aperture
(\TUBE{}) that decodes directly to a SystemVerilog testbench monitor. When
firmware writes to the \TUBE{}, the monitor accumulates characters until a
newline and emits the message into the UVM log. Because severity indicators
(info, warning, error, fatal) trigger immediate UVM scoreboard checks and
end-of-test (EOT) objection handling, log messages must materialize inside the
simulator at the exact simulated timestamp they are emitted.

\subsection{Anatomy of the Simulation Logging Tax}
\label{sec:anatomy}

Even with a zero-baud-delay \TUBE{}, legacy firmware logging compounds overhead
across three architectural layers:

\paragraph{Software Formatting on the Simulated CPU Pipeline}
Converting integer and hexadecimal arguments into ASCII requires iterative
division/modulo loops, specifier parsing, and buffer copying that the RTL
simulator must evaluate cycle-by-cycle across fetch, decode, ALU, and retirement
stages. Eliminating on-core formatting and byte-store loops reduces per-call
retired CPU instructions by $69.6\%\text{--}70.4\%$ ($838 \to 255$ and
$874 \to 259$, Table~\ref{tbl:trace_decomp}) and shrinks full-chip instruction
traces across \SUITEA{} by $42.84\%$ median (up to $79.78\%$).

\paragraph{Bus Interconnect Serialization}
Across the three subsystems we study, two write \emph{one character per
transaction} through a byte-wide aperture ($55$ interconnect transactions for a
$55$-character message), while the third packs four format-string characters per
$32$-bit write plus framing and argument words ($20$ transactions). Every write
traverses CPU store buffers, interconnect arbiters, crossbars, and bridge
slices, triggering thousands of RTL logic evaluation events.

\paragraph{Memory Ordering Barriers}
One subsystem's register-access wrapper pairs \emph{every} character store with
an explicit system data-synchronization barrier, draining the
store buffer $55$ times per message and preventing write coalescing. The other
two issue volatile stores ordered by the aperture's strongly ordered, non-gathering device-memory
attributes, which stall once the processor's $3$-entry
MMIO store buffer fills (Section~\ref{sec:costmodel}).

\paragraph{Compounding Disk and Host I/O Overhead}
Formatting instructions flood retirement traces, MMIO bursts generate dense
interconnect waveform toggles, and repeated bus handshakes fill transaction
logs. Collapsing each print to $(1+\nslots)$ word writes shrinks instruction
traces by $42.84\%$ median, bus logs by $38.17\%$ median, and compressed
waveform databases by $28.82\%$ median ($2.98\text{--}5.94\,\text{GB}$ saved per
deep debug run; Section~\ref{sec:eval}).

\section{Cost Model}
\label{sec:costmodel}

To quantify why host-side log tokenization yields substantial simulation speedups, we derive a closed-form analytical cost model that relates per-message instruction and bus-transaction reductions to end-to-end simulation speedup and effective simulator throughput.

\subsection{Per-message cost}

Consider a single log message of formatted length $\msglen$ characters taking $\nargs$ format arguments. A legacy print costs an on-core formatting loop, some number of ordered writes to the \TUBE{} aperture, and, on subsystems whose driver requires it, an ordering barrier per write:
\begin{equation}
  \Clegacy = \Cfmt(\msglen) + N_{\mathrm{wr}}(\msglen)\,\Cwr + N_{\mathrm{bar}}(\msglen)\,\Cbar,
  \label{eq:clegacy}
\end{equation}
where $\Cfmt(\msglen)$ is the instruction cost of the on-core formatting routine, $\Cwr$ is the topology-dependent cost of a single ordered write transaction from the initiating core to its \TUBE{} aperture, and $\Cbar$ is the cost of an ordering barrier.

There is deliberately no status-poll term: every deployed print path in this SoC is fire-and-forget, and the testbench receivers are passive bus-clock monitors with no wait-state counter, ready/valid backpressure, or DPI callback. Thus $\Cwr$ is purely initiator-side store issue and interconnect traversal rather than a testbench handshake. The coefficients depend on each subsystem's print driver and interconnect path, and the three drivers deployed across our full-chip verification environment directly explain the empirical suite deltas in Section~\ref{sec:eval}:
\begin{itemize}
  \item \textbf{Byte-serial with on-core formatting and per-character barriers (\SUITEA{} primary core):} C library \texttt{vsprintf} renders into a stack buffer ($\Cfmt = 1{,}873$ cycles, $612$ retired instructions for a $30$-char, $1$-arg call; Table~\ref{tbl:trace_decomp}), then issues one store and one full system data-synchronization barrier per character ($N_{\mathrm{wr}} = N_{\mathrm{bar}} = \msglen$).
  \item \textbf{Byte-serial with on-core formatting (shared application cores):} Local \texttt{vsnprintf} formats on-core ($\Cfmt = 3{,}677$ cycles [$1{,}833.5\,\text{ns}$], $418$ instructions; we use \texttt{vsprintf} as unified shorthand for on-core radix formatting $\Cfmt$), followed by one single-byte cross-fabric store per character without an explicit barrier ($N_{\mathrm{wr}} = \msglen$, $N_{\mathrm{bar}} = 0$).
  \item \textbf{Word-packed format-string shipping with host-side formatting (\SUITEB{} primary core):} Performs no on-core radix conversion ($\Cfmt = 0$), but runs a byte-packing loop to ship the raw format string four characters per 32-bit MMIO write alongside framing and raw argument words ($N_{\mathrm{wr}} = \lceil \msglen/4 \rceil + \nargs + 4$, $N_{\mathrm{bar}} = 0$) while the SystemVerilog monitor formats on the host.
\end{itemize}
Under tokenization all three drivers collapse to a single token word plus $\nslots$ argument words, zero formatting or byte packing ($\Cfmt = 0$), and at most one trailing barrier per message:
\begin{equation}
  \Ctoken = \Cpush(\nargs) + (1 + \nslots)\,\Cwr + b\,\Cbar, \qquad b \in \{0,1\},
  \label{eq:ctoken}
\end{equation}
where $\nslots \ge \nargs$ ($1$ slot per 32-bit scalar, $2$ per 64-bit scalar, $\lceil (\ell+1)/4 \rceil$ for an $\ell$-byte runtime string; Section~\ref{sec:design}) and $\Cpush(\nargs)$ is the single-pass specifier scan that stores promoted register values to the aperture with zero radix conversion or division.

\begin{figure}[!t]
\centering
\begin{tikzpicture}[font=\scriptsize, x=0.295cm, y=0.35cm]
  \draw[gray!40, very thin] (0,-0.4) -- (0,5.8);
  \foreach \x/\lbl in {0/0, 3/3, 6/6, 9/9, 12/12, 15/15} {
    \draw[gray!30, densely dotted] (\x,-0.4) -- (\x,5.6);
    \node[below, font=\tiny] at (\x,-0.4) {\lbl};
  }
  \node[below, font=\scriptsize] at (7.5,-1.05) {Single-Call On-Core Execution Time ($\mu\text{s}$, $\msglen{=}30, \nslots{=}1$)};

  \node[left, font=\tiny] at (-0.1,5.2) {\textbf{Cross-Fab.\ Base}};
  \fill[gray!25, draw=black] (0,4.9) rectangle (0.500,5.5);
  \fill[gray!55, draw=black] (0.500,4.9) rectangle (2.334,5.5);
  \fill[gray!85, draw=black] (2.334,4.9) rectangle (15.234,5.5);
  \node[right, font=\tiny] at (15.32,5.2) {$15.23\,\mu\text{s}$ ($84.7\%$ bus)};

  \node[left, font=\tiny] at (-0.1,4.3) {\textbf{Cross-Fab.\ Tok}};
  \fill[gray!25, draw=black] (0,4.0) rectangle (0.522,4.6);
  \fill[gray!85, draw=black] (0.522,4.0) rectangle (1.382,4.6);
  \node[right, font=\tiny] at (1.50,4.3) {$\mathbf{1.38\,\mu\text{s}}$ ($\mathbf{-90.9\%}$)};

  \node[left, font=\tiny] at (-0.1,3.1) {\textbf{Local Tube Base}};
  \fill[gray!25, draw=black] (0,2.8) rectangle (0.105,3.4);
  \fill[gray!55, draw=black] (0.105,2.8) rectangle (1.978,3.4);
  \fill[gray!85, draw=black] (1.978,2.8) rectangle (3.037,3.4);
  \node[right, font=\tiny] at (3.15,3.1) {$3.04\,\mu\text{s}$ ($61.7\%$ \texttt{vsprintf})};

  \node[left, font=\tiny] at (-0.1,2.2) {\textbf{Local Tube Tok}};
  \fill[gray!25, draw=black] (0,1.9) rectangle (1.039,2.5);
  \fill[gray!85, draw=black] (1.039,1.9) rectangle (1.110,2.5);
  \node[right, font=\tiny] at (1.25,2.2) {$\mathbf{1.11\,\mu\text{s}}$ ($\mathbf{-63.5\%}$, $-93.3\%$ bus)};

  \node[left, font=\tiny] at (-0.1,1.0) {\textbf{Word-Pack Base}};
  \fill[gray!25, draw=black] (0,0.7) rectangle (0.105,1.3);
  \fill[gray!85, draw=black] (0.105,0.7) rectangle (3.058,1.3);
  \node[right, font=\tiny] at (3.17,1.0) {$3.06\,\mu\text{s}$ ($13$ packed writes)};

  \node[left, font=\tiny] at (-0.1,0.1) {\textbf{Word-Pack Tok}};
  \fill[gray!25, draw=black] (0,-0.2) rectangle (0.622,0.4);
  \fill[gray!85, draw=black] (0.622,-0.2) rectangle (1.076,0.4);
  \node[right, font=\tiny] at (1.20,0.1) {$\mathbf{\approx 1.08\,\mu\text{s}}$ ($\mathbf{-64.8\%}$, $-84.6\%$ bus)};

  \fill[gray!25, draw=black] (-3.8,6.05) rectangle (-3.1,6.45);
  \node[right, font=\tiny] at (-3.0,6.25) {Prolog / Scan ($\Cpush$)};
  \fill[gray!55, draw=black] (4.8,6.05) rectangle (5.5,6.45);
  \node[right, font=\tiny] at (5.6,6.25) {\texttt{vsprintf} ($\Cfmt$)};
  \fill[gray!85, draw=black] (12.2,6.05) rectangle (12.9,6.45);
  \node[right, font=\tiny] at (13.0,6.25) {MMIO Bus + Barrier};
\end{tikzpicture}
\vspace{-4pt}
\caption{Two-regime single-call cost decomposition ($\msglen=30, \nslots=1$, Table~\ref{tbl:trace_decomp}). The light gray segment denotes prolog/epilog alone in baseline bars ($\Cpush=0$) and prolog/epilog plus specifier scan ($\Cpush$) in tokenized bars. Cross-fabric logging is dominated by MMIO bus stalls ($84.7\%$), whereas local subsystem logging is dominated by on-core \texttt{vsprintf} ($61.7\%$). Tokenization removes $\Cfmt$ and cuts MMIO bus-stall cycles by $84.6\%\text{--}93.3\%$.}
\label{fig:trace_regimes}
\end{figure}
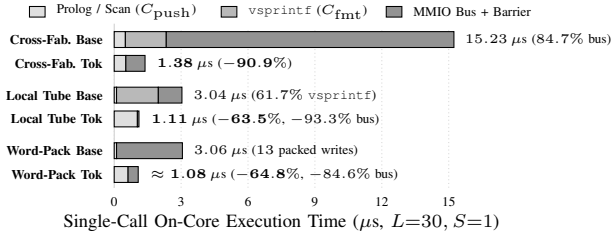

\paragraph{Hardware trace and timing decomposition across all three interconnect topologies}
To parameterize Equation~\eqref{eq:clegacy} without fitting to suite averages, we extract cycle- and instruction-exact parameters from processor instruction-retirement traces for the two processor-formatted topologies (Items 1--2) and from the hardware-timestamped CSR race window for the word-packed controller (Item 3, where instruction tracing is disabled in regression; Section~\ref{sec:latent}). Four message instances appear in our analysis: $\msglen=30$ ($\nslots=1$, single-call reference in Table~\ref{tbl:trace_decomp} and Figure~\ref{fig:trace_regimes}), $\msglen=34$ ($\nslots=4$, multi-argument trace call), $\msglen=38$ ($\nslots=1$, CSR race print in Section~\ref{sec:latent}), and $\msglen=55$ ($\nslots=2$, bus-transaction comparison):
\begin{enumerate}
  \item \textbf{Cross-fabric SoC tube (64-bit application core at $2.0\,\text{GHz}$ nominal, $N_{\mathrm{wr}}=\msglen, N_{\mathrm{bar}}=0$):} For $\msglen=30, \nslots=1$, the baseline print routine takes $15{,}234.0\,\text{ns}$ ($30{,}553$ cycles, $838$ instructions): on-core formatting is $\Cfmt = 1{,}833.5\,\text{ns}$ ($3{,}677$ cycles, $418$ instructions, $12.0\%$), while $30$ single-byte stores across the cross-fabric peripheral bridge take $\msglen\cdot\Cwr = 12{,}900.1\,\text{ns}$ ($25{,}872$ cycles, $420$ instructions, $84.7\%$, or $430.0\,\text{ns} = 862.4$ cycles/char). A $3$-character string completes the byte-write loop in $0.997\,\text{ns/char}$ ($2.0$ cycles/char) by absorbing writes into the processor's $3$-entry non-gathering MMIO store buffer, after which every subsequent byte store stalls on the cross-fabric handshake ($422\text{--}434\,\text{ns/char}$). Under tokenization ($1+\nslots=2$ stores), in the pessimistic steady-state back-to-back logging regime where the $3$-entry buffer is already occupied ($2 \times 430.0 = 860.0\,\text{ns}$ bus stall, Table~\ref{tbl:trace_decomp}), the call completes in $1{,}381.6\,\text{ns}$ ($2{,}771$ cycles, $255$ instructions), an \textbf{$11.03\times$ ($-90.93\%$)} cycle reduction saving $13.85\,\mu\text{s}$---whereas an isolated call starting from an empty store buffer absorbs both stores in $2.0\,\text{ns}$ and completes in $\approx 524\,\text{ns}$ (\textbf{$29.1\times$}, so our $11.03\times$ headline is a conservative lower bound). On a $34$-character, $4$-argument call ($\msglen=34, \nslots=4$, $1+\nslots=5$ stores; baseline $19{,}595.3\,\text{ns}$ [$500.4\,\text{ns}$ prolog, $\Cfmt = 4{,}474.9\,\text{ns}$, $34 \times 430.0 = 14{,}620.0\,\text{ns}$ bus]), a cold buffer absorbs the first $3$ stores in $3.0\,\text{ns}$ so only stores $4$--$5$ stall ($2 \times 430.0 = 860.0\,\text{ns}$, plus $15.0\,\text{ns}$ prolog and $\Cpush(4) = 1{,}227.1\,\text{ns}$ [$506.6\,\text{ns}$ base $+ 3 \times 240.2\,\text{ns}$]), completing in $2{,}105.1\,\text{ns}$ ($9.31\times$, saving $17.49\,\mu\text{s}$); with a full buffer ($5 \times 430.0 = 2{,}150.0\,\text{ns}$ bus), it completes in $3{,}392.1\,\text{ns}$ ($5.78\times$, $-82.7\%$, saving $16.20\,\mu\text{s}$).
  \item \textbf{Subsystem-local tube (\SUITEA{} primary 64-bit core at $1.000\,\text{GHz}$, $1{,}000\,\text{ps/cycle}$, $N_{\mathrm{wr}}=N_{\mathrm{bar}}=\msglen$):} For a local subsystem aperture, the steady-state per-character store-plus-barrier round-trip ($\Cwr + \Cbar$) is $27.0\,\text{cycles}$ ($27.0\,\text{ns}$: $19.0\,\text{cycles}$ forward traversal, $7.0\,\text{cycles}$ write response, $1.0\,\text{cycle}$ retirement, or $35.3\,\text{cycles/char}$ with $8.3\,\text{cycles}$ loop branch overhead). For $\msglen=30, \nslots=1$ ($3{,}037.0\,\text{ns}$, $3{,}037$ cycles, $874$ instructions), \texttt{vsprintf} ($\Cfmt = 1{,}873$ cycles, $612$ instructions) dominates \textbf{$61.7\%$} of baseline latency while the $30$ barriered writes ($\msglen(\Cwr+\Cbar) = 1{,}059$ cycles, $180$ instructions) take \textbf{$34.9\%$}. Tokenization eliminates \texttt{vsprintf} ($\Cfmt = 0$), replacing the $30$ barriered byte writes with $\Cpush = 923$ cycles ($218$ instructions), $2$ back-to-back posted MMIO stores ($28$ cycles, $14$ cycles/store), and one trailing data-synchronization barrier ($\Cbar = 43$ cycles to drain two word stores from a cold barrier state vs.\ $27.0$ steady-state cycles per byte store+barrier in a warm loop, cutting bus-and-barrier cycles from $1{,}059$ to $28+43=71$, or $-93.3\%$). Total call time drops to $1{,}110.0\,\text{ns}$ ($259$ instructions)—a \textbf{$2.74\times$ ($-63.45\%$ cycle, $-70.37\%$ instruction)} reduction (\textbf{$3.45\times$ / $-70.99\%$} on $0$-argument prints).
  \item \textbf{Word-packed subsystem tube (\SUITEB{} primary controller at $1.000\,\text{GHz}$-equivalent, $N_{\mathrm{wr}}=\lceil\msglen/4\rceil+\nargs+4, N_{\mathrm{bar}}=0$):} In the hardware-timestamped CSR race window of Section~\ref{sec:latent}, a $38$-character, $1$-argument call issues $N_{\mathrm{wr}} = \lceil 38/4 \rceil + 1 + 4 = 15$ ordered 32-bit MMIO writes in $3{,}407\,\text{ns}$ ($\Cwr \approx 227.1\,\text{ns/tx}$ including the byte-packing loop); tokenization collapses those $15$ writes to $2$ ($\approx 454\,\text{ns}$), saving $\approx 2.95\,\mu\text{s}$ ($-86.7\%$). At the $\msglen=30, \nslots=1$ reference point ($N_{\mathrm{wr}} = \lceil 30/4 \rceil + 1 + 4 = 13$ writes), baseline MMIO serialization takes $13 \times 227.1 = 2{,}952.7\,\text{ns}$ versus $2 \times 227.1 = 454.3\,\text{ns}$ tokenized ($-84.6\%$ of MMIO bus time; Table~\ref{tbl:trace_decomp}).
\end{enumerate}

\begin{table*}[!t]
\caption{Single-call timing decomposition of a $30$-character, $1$-argument \texttt{printf} ($\msglen{=}30, \nslots{=}1$) across all three deployed SoC logging topologies}
\label{tbl:trace_decomp}
\centering
\footnotesize
\setlength{\tabcolsep}{5.5pt}
\begin{tabular}{@{}lrrrrr@{}}
\toprule
 & \multicolumn{2}{c}{\textbf{Cross-Fabric Tube}} & \multicolumn{2}{c}{\textbf{Subsystem-Local Tube}} & \textbf{Word-Packed Tube$^{\S}$} \\
 & \multicolumn{2}{c}{\textbf{(App.\ Core, $2.0\,\text{GHz}$ nom.)}} & \multicolumn{2}{c}{\textbf{(\SUITEA{}, $1.000\,\text{GHz}$)}} & \textbf{(\SUITEB{}, $1.000\,\text{GHz}$)} \\
\cmidrule(lr){2-3} \cmidrule(lr){4-5} \cmidrule(l){6-6}
\textbf{Phase / Cost Term} & \textbf{Time (ns)} & \textbf{Retired Inst.} & \textbf{Time (ns)} & \textbf{Retired Inst.} & \textbf{Time (ns)} \\
\midrule
\textbf{Baseline Total ($\Clegacy$)} & \textbf{15,234.0} & \textbf{838} & \textbf{3,037.0} & \textbf{874} & \textbf{$\approx$3,058 (2,952.7 bus)} \\
\quad Prolog / Epilog & 500.4 (3.3\%) & 0$^{\dagger}$ & 105.0 (3.5\%) & 82 & $\approx$105 (ref.) \\
\quad $\Cfmt$ (on-core \texttt{vsprintf}) & 1,833.5 (12.0\%) & 418 & 1,873.0 (61.7\%) & 612 & 0.0 (0.0\%) \\
\quad $N_{\mathrm{wr}}(\Cwr{+}\Cbar)$ (MMIO + barrier)$^{\ddagger}$ & 12,900.1 (84.7\%) & 420 & 1,059.0 (34.9\%) & 180 & 2,952.7 (96.6\%) \\
\quad \emph{Effective rate per bus transaction} & \emph{430.0 ns/tx} & \emph{14.0 inst/tx} & \emph{35.3 ns/tx} & \emph{6.0 inst/tx} & \emph{227.1 ns/tx} \\
\midrule
\textbf{Tokenized Total ($\Ctoken$)} & \textbf{1,381.6} & \textbf{255} & \textbf{1,110.0} & \textbf{259} & \textbf{$\approx$1,076 (454.3 bus)} \\
\quad Prolog / Epilog & 15.0 (1.1\%) & 34 & 116.0 (10.5\%) & 38 & $\approx$116 (ref.) \\
\quad $\Cpush(\nargs)$ (specifier scan) & 506.6 (36.7\%) & 219 & 923.0 (83.2\%) & 218 & $\approx$506 (ref.) \\
\quad $(1{+}\nslots)\Cwr{+}b\Cbar$ (MMIO + barrier)$^{\ddagger}$ & 860.0 (62.2\%) & 2 & 71.0 (6.4\%) & 3 & 454.3 (42.2\%) \\
\midrule
\textbf{Net Per-Call Savings ($\Delta C$)} & \textbf{-13,852.4} & \textbf{-583} & \textbf{-1,927.0} & \textbf{-615} & \textbf{$\approx$-1,982 (-2,498.4 bus)} \\
\textbf{Relative Reduction (Speedup)} & \textbf{-90.9\% (11.03$\times$)} & \textbf{-69.6\%} & \textbf{-63.5\% (2.74$\times$)} & \textbf{-70.4\%} & \textbf{$\approx$-64.8\% (2.84$\times$; -84.6\% bus)} \\
\bottomrule
\end{tabular}

\vspace{2pt}
\parbox{\textwidth}{\scriptsize $^{\dagger}$Included in formatting frame setup. $^{\ddagger}$On Cross-Fabric ($N_{\mathrm{wr}}{=}30$) and Word-Packed ($N_{\mathrm{wr}}{=}13$), $N_{\mathrm{bar}}{=}0$ ($\Cbar{=}0$); on Local ($N_{\mathrm{wr}}{=}30$), $N_{\mathrm{bar}}{=}30$ baseline and $b{=}1$ tokenized. The Cross-Fabric tokenized bus term ($2 \times 430.0 = 860.0\,\text{ns}$) reflects steady-state back-to-back logging with the $3$-entry non-gathering MMIO store buffer occupied; when the buffer starts empty, $1\text{--}3$ stores are absorbed in $2.0\text{--}3.0\,\text{ns}$ (reducing $\msglen{=}30, \nslots{=}1$ to $\approx 524\,\text{ns}$ [$29.1\times$], and $\msglen{=}34, \nslots{=}4$ [$5$ stores] from $3{,}392.1\,\text{ns}$ [$5.78\times$ full-buffer] to $2{,}105.1\,\text{ns}$ [$9.31\times$ cold-buffer]). $^{\S}$In the Word-Packed column (where instruction tracing is disabled in regression), only $\Cwr \approx 227.1\,\text{ns/tx}$ ($3{,}407\,\text{ns}/15$ tx in Section~\ref{sec:latent}) is directly measured ($2{,}952.7 \to 454.3\,\text{ns}$, $-84.6\%$ bus time; $-86.7\%$ at $\msglen{=}38$), whereas prolog/epilog ($\approx 105\text{ / }116\,\text{ns}$) and $\Cpush$ ($\approx 506\,\text{ns}$) are composite reference estimates from the adjacent columns (substituting the $1.0\,\text{GHz}$ local $\Cpush{=}923\,\text{ns}$ yields $\approx 1{,}493\,\text{ns}$ tokenized total, or $-51.2\%$ [$2.05\times$] total and $-84.6\%$ bus reduction).}
\end{table*}

Table~\ref{tbl:trace_decomp} and Figure~\ref{fig:trace_regimes} show that both software formatting ($\Cfmt$) and bus serialization ($\Cwr + \Cbar$) are first-order bottlenecks, with interconnect distance determining which term dominates: on local subsystem tubes ($\Cwr+\Cbar = 27\,\text{ns}$), \texttt{vsprintf} formatting ($\Cfmt = 1{,}873\,\text{ns}$) is $61.7\%$ of the call cost; on cross-fabric SoC tubes ($\Cwr = 430\,\text{ns}$ once the $3$-entry MMIO store buffer fills), per-character bus serialization ($\msglen\cdot\Cwr = 12{,}900\,\text{ns}$) is $84.7\%$ of the cost. In full-chip simulation---combining cross-fabric application-core prints ($\approx 13.9\text{--}17.5\,\mu\text{s}$ saved) with subsystem-local prints ($\approx 1.9\text{--}3.0\,\mu\text{s}$ saved)---critical-path savings average $15.27\,\mu\text{s}$ unweighted ($15.08\,\mu\text{s}$ print-weighted) on \SUITEA{} ($15.21\,\mu\text{s}$ / $15.06\,\mu\text{s}$ across all eight multi-core workloads) and $3.64\text{--}5.41\,\mu\text{s}$ on peripheral-dominated \SUITEB{} workloads (Section~\ref{sec:eval}).

\paragraph{Collapse of interconnect transactions to $1+\nslots$}
For a $55$-character message with two integer arguments ($\nslots = 2$), tokenization reduces interconnect traffic to $3$ word writes: $55 \rightarrow 3$ ($-94.5\%$, $18.3\times$) against byte-serial transport, and $20 \rightarrow 3$ ($-85.0\%$, $6.7\times$) against word-packed transport ($\lceil 55/4 \rceil + 2 + 4 = 20$ writes). Unlike embedded silicon logging~\cite{pwtokenizer,nanolog2018}, where minimizing wire bytes via variable-length integer encoding is paramount, RTL simulation pays a fixed interconnect evaluation cost per bus transaction regardless of payload bit-width. A fixed 32-bit word-aligned slot protocol matches this cost function directly, minimizing bus transactions with zero bit-shifting or packing overhead on the simulated CPU.

\subsection{From per-message savings to end-to-end speedup and throughput (\texorpdfstring{$R_{\mathrm{eff}}$}{R\_eff})}

Let a simulation workload execute $N$ log messages over a baseline simulated cycle count $T_{\mathrm{legacy}}$, with baseline \emph{print density} $\density \triangleq N / T_{\mathrm{legacy}}$. Because tokenization preserves non-logging instruction sequences identically (verified by SHA-256 binary parity in Section~\ref{sec:eval}), in workloads without timing-sensitive multi-core races (Section~\ref{sec:latent}) the tokenized cycle count is $T_{\mathrm{token}} = T_{\mathrm{legacy}} - N \cdot \Delta C$, where $\Delta C \triangleq \Clegacy - \Ctoken$ is the workload-weighted per-message cycle saving. Taking the serial-removal limit ($s \to \infty$) of Amdahl's law~\cite{amdahl1967} yields the simulated-time speedup and its linear reciprocal:
\begin{equation}
  \speedup(\density) = \frac{T_{\mathrm{legacy}}}{T_{\mathrm{token}}} = \frac{1}{1 - \density \, \Delta C}
  \quad\Longleftrightarrow\quad
  \frac{1}{\speedup} = 1 - \Delta C \cdot \density.
  \label{eq:speedup}
\end{equation}
Across our production SoC suites (Section~\ref{sec:eval}), simulated cycle reduction increases monotonically with print volume up to $-78.44\%$ ($4.639\times$ cycle speedup), providing an algebraic consistency check on Equation~\eqref{eq:speedup} (where per-test print counts $\density$ are instrumented on our deep-dive subset rather than across all $858$ pairs).

Host simulator CPU time falls even further than simulated cycles across both suites: by $-41.51\%$ versus $-33.32\%$ simulated time ($+8.19$ points) on \SUITEA{} ($N=169$), and by $-21.33\%$ CPU / $-22.10\%$ wall time versus $-12.43\%$ simulated time ($+8.90\text{--}9.67$ points) on \SUITEB{} ($N=689$). Rather than claiming a super-linear simulator mechanism, we note that host CPU time is the product of simulated cycle count and average per-cycle evaluation cost, and tokenization reduces both because logging cycles are more expensive to evaluate than average. Defining \emph{effective simulator throughput} and its baseline-normalized ratio for pair $i$ as
\begin{equation}
  R_{\mathrm{eff},i} \triangleq \frac{T_{\mathrm{sim},i}}{T_{\mathrm{CPU},i}}, \qquad
  \widehat{R}_{\mathrm{eff},i} \triangleq \frac{R_{\mathrm{eff,tok},i}}{R_{\mathrm{eff,base},i}},
  \label{eq:hzeff}
\end{equation}
where $\widehat{R}_{\mathrm{eff,base},i} \equiv 1.000\times$ by construction and $S_{\mathrm{CPU},i} \triangleq T_{\mathrm{CPU,base},i}/T_{\mathrm{CPU,tok},i} = \speedup_i \cdot \widehat{R}_{\mathrm{eff},i}$, $R_{\mathrm{eff}}$ is a derived ratio of two measured quantities reporting the change in average per-cycle evaluation cost (driven by high-fanout interconnect handshakes, store-buffer drains, instruction-trace writes, and waveform toggles during logging windows). Excising these windows raises $\widehat{R}_{\mathrm{eff}}$ across \SUITEA{} by $+24.02\%$ mean per CPU-s ($1.177\times$ GMean, up to $+147.63\%$; $+8.51\%$ mean, $1.043\times$ GMean, up to $+113.87\%$ per wall-s) and across \SUITEB{} by $+19.60\%$ mean per CPU-s ($1.113\times$ GMean, up to $+211.96\%$; $+23.25\%$ mean, $1.145\times$ GMean, up to $+262.21\%$ per wall-s).

\section{Related Work}
\label{sec:related}

\textbf{Deferred and tokenized software logging.}
Replacing format strings with compact identifiers is standard in embedded and server logging (\texttt{pw\_tokenizer}~\cite{pwtokenizer}, \texttt{defmt}~\cite{defmt}, \texttt{NanoLog}~\cite{nanolog2018}, Zephyr dictionary logging~\cite{zephyrdict}, \texttt{TRICE}~\cite{trice}, Memfault Compact Logs~\cite{memfaultcompact}, and OS tracers such as DTrace~\cite{dtrace} and LTTng~\cite{lttng} that emit structured records~\cite{consolelogs}). However, embedded tokenizers optimize for \emph{wire bytes and flash footprint} via variable-length integer packing, require call-site macro rewrites, and decode \emph{offline} after execution completes. In RTL simulation, variable-length byte packing adds CPU shift/mask instructions without saving interconnect transactions ($\Cwr$), and offline decoding breaks live UVM scoreboards~\cite{uvm} that inspect logs dynamically at each simulation timestamp ($\Delta t_{\mathrm{sim}} = 0$).

\textbf{Simulation I/O, trace IP, and simulator acceleration.}
Within pre-silicon verification, behavioral UART FIFOs eliminate baud-rate delays but still pay on-core formatting ($\Cfmt$) and per-character MMIO stalls. Arm Semihosting~\cite{armsemihosting}, SystemC TLM-2.0 virtual platforms~\cite{ieee1666}, and DPI-C memory backdoors bypass architectural load/store execution and interconnect fabrics, whereas hardware trace transports (MIPI STP~\cite{mipistp}, CoreSight STM~\cite{coresightstm}) and FPGA trace synthesis (FireSim~\cite{firesim2018}) require dedicated trace RTL or out-of-band bridges (violating \textbf{R5}) and decode offline. Similarly, RTL simulator kernel optimizations—such as compiled C++ evaluation in Verilator~\cite{verilator}, activity-skipping in ESSENT~\cite{essent}, and parallel partitioning in RepCut~\cite{repcut}—accelerate per-cycle evaluation speed, whereas our approach removes millions of target logging cycles and bus handshakes from the workload itself. Unlike host-side UVM \texttt{uvm\_info} verbosity filtering, our architecture eliminates on-target firmware formatting and interconnect serialization while preserving $100\%$ live log visibility (Table~\ref{tbl:related}).

\begin{table}[t]
\caption{Comparison of logging architectures across target domains and verification requirements}
\label{tbl:related}
\centering
\scriptsize
\setlength{\tabcolsep}{1.3pt}
\begin{tabular}{@{}lccccc@{}}
\toprule
\textbf{System} & \textbf{Domain} & \textbf{Metric} & \textbf{Token} & \textbf{Decode} & \textbf{Edits} \\
\midrule
pw\_tokenizer~\cite{pwtokenizer} & MCU & Flash/Wire & Hash & Offline & Macro \\
defmt~\cite{defmt} & MCU/Rust & Flash/Wire & ELF Idx & Offline & Macro \\
NanoLog~\cite{nanolog2018} & Server & Host Lat. & Static ID & Offline & Macro \\
Zephyr~\cite{zephyrdict}/TRICE~\cite{trice} & MCU/RTOS & Wire Bytes & ID/Idx & Offline & Macro \\
DPI Backdoor & Core TB & Sim.\ Cyc. & Mem.\ VA & In-Sim & Hook \\
\textbf{This Work} & \textbf{SoC RTL} & \textbf{Txns/Cyc.} & \textbf{ELF VA} & \textbf{Live ($\Delta t{=}0$)} & \textbf{None} \\
\bottomrule
\end{tabular}
\end{table}

\section{Design}
\label{sec:design}

\subsection{Architectural requirements}

Our design is governed by five verification requirements:
\begin{itemize}
  \item \textbf{R1 --- Bit-Exact Output:} Reconstructed log streams must be byte-for-byte identical to legacy character output so UVM scoreboards and golden diffs pass without changes.
  \item \textbf{R2 --- Zero Call-Site Edits:} Firmware engineers must not modify existing \texttt{printf} call sites across legacy source files.
  \item \textbf{R3 --- Strictly Opt-In Build Parity:} When tokenization is disabled, compiled binaries and simulation behavior must be 100\% byte-identical to the baseline build.
  \item \textbf{R4 --- ISA-Independent Multi-Architecture Decoding:} A single testbench decoder must service heterogeneous 32-bit and 64-bit cores across Arm (\texttt{AArch64}/\texttt{AArch32}), RISC-V (\texttt{RV32}/\texttt{RV64}), and DSP subsystems with differing ABI conventions.
  \item \textbf{R5 --- Zero Hardware or Interconnect Changes:} The protocol must reuse existing memory-mapped simulation tube apertures without adding RTL logic or modifying the design under test (DUT).
\end{itemize}

\subsection{Token identity: link-time ELF virtual address}
\label{sec:tokenid}

Compile-time string-hashing tokenizers~\cite{pwtokenizer} require a preprocessor macro to inspect string literals at every call site, violating \textbf{R2}; once execution enters a variadic C function, the literal has decayed to a \texttt{const char *} pointer. In linked firmware images, however, the \texttt{const char *} pointer passed to \texttt{tok\_printf(const char *fmt, ...)} \emph{is} already a unique identifier: its link-time virtual address in read-only data (\texttt{.rodata}). Using the 32-bit ELF virtual address as the token yields three properties:
\begin{enumerate}
  \item \textbf{Zero call-site edits (\textbf{R2}):} Redirecting a subsystem's logging requires only a single header alias mapping its print macro to \texttt{tok\_printf}.
  \item \textbf{Collision-free across distinct strings:} Within a firmware image, two format strings share a \texttt{.rodata} address only if the linker merges textually identical or suffix-overlapping literals (\texttt{-fmerge-constants}); because the dictionary string is identical, reconstructed output remains bit-exact (whereas hash tokenizers also admit collisions between \emph{distinct} strings). Across subsystems with overlapping local SRAM/TCM address maps, each simulation tube aperture binds exclusively to its core's sidecar dictionary indexed by $(\text{aperture\_id}, \text{VA})$.
  \item \textbf{CPU architecture independence (\textbf{R4}):} Unlike Arm Semihosting~\cite{armsemihosting} (\texttt{BKPT}/\texttt{HLT} traps), ELF \texttt{.rodata} sections are standard across \texttt{gcc} and \texttt{clang}/\texttt{LLVM} toolchains for Arm, RISC-V, and DSP targets.
\end{enumerate}

\subsection{Word-aligned slot wire protocol and severity handling}
\label{sec:wire}

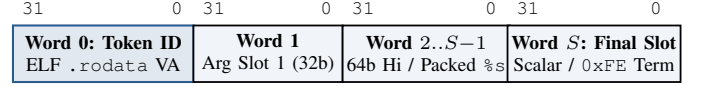
\begin{figure}[t]
\centering
\resizebox{\columnwidth}{!}{%
\begin{tikzpicture}[
  font=\rmfamily\scriptsize,
  reg/.style={draw=black, line width=0.65pt, fill=pgray, minimum height=7.2mm, align=center, inner sep=1.5pt},
  reghdr/.style={draw=pblue, line width=0.75pt, fill=pblue!14, minimum height=7.2mm, align=center, inner sep=1.5pt}
]
\node[font=\ttfamily\scriptsize, anchor=south west] at (-4.20,1.08) {31};
\node[font=\ttfamily\scriptsize, anchor=south east] at (-1.90,1.08) {0};
\node[font=\ttfamily\scriptsize, anchor=south west] at (-1.90,1.08) {31};
\node[font=\ttfamily\scriptsize, anchor=south east] at (0.00,1.08) {0};
\node[font=\ttfamily\scriptsize, anchor=south west] at (0.00,1.08) {31};
\node[font=\ttfamily\scriptsize, anchor=south east] at (2.10,1.08) {0};
\node[font=\ttfamily\scriptsize, anchor=south west] at (2.10,1.08) {31};
\node[font=\ttfamily\scriptsize, anchor=south east] at (4.20,1.08) {0};

\node[reghdr, minimum width=2.30cm, anchor=west] (w0) at (-4.20,0.70)
  {\textbf{Word 0: Token ID}\\{\scriptsize ELF \texttt{.rodata} VA}};
\node[reg, minimum width=1.90cm, anchor=west] (w1) at (-1.90,0.70)
  {\textbf{Word 1}\\{\scriptsize Arg Slot 1 (32b)}};
\node[reg, minimum width=2.10cm, anchor=west] (w2) at (0.00,0.70)
  {\textbf{Word $2..\nslots{-}1$}\\{\scriptsize 64b Hi / Packed \texttt{\%s}}};
\node[reg, minimum width=2.10cm, anchor=west] (ws) at (2.10,0.70)
  {\textbf{Word $\nslots$: Final Slot}\\{\scriptsize Scalar / \texttt{0xFE} Term}};
\end{tikzpicture}%
}
\caption{Word-aligned 32-bit MMIO wire packet layout ($1+\nslots$ bus writes on the target processor, reconstructed in $\Delta t_{\mathrm{sim}} = 0$ by the SystemVerilog monitor).}
\label{fig:decoder_flow}
\end{figure}

A tokenized log message is transmitted as a 32-bit token word followed by $\nslots$ 32-bit argument slots (Table~\ref{tbl:slots}, Figure~\ref{fig:decoder_flow}). Promoting narrow integers to 32-bit slots and serializing 64-bit values as two deterministic 32-bit words (low word first, then high word) satisfies \textbf{R4} across \texttt{LP64}, \texttt{ILP32}, and DSP ABIs. Dynamic runtime strings (\texttt{\%s}) pack four characters per 32-bit word terminated by a \texttt{0xFE} sentinel. To preserve exact UVM error-counter timing (\textbf{R1}), warning/error prints retain a single control-byte write ahead of the token word ($2+\nslots$ writes on $<2\%$ of messages in passing runs). Post-link Python extraction parses each ELF image's \texttt{.rodata} section into a sidecar dictionary (\DICTFILENAME{}) recording each string's 32-bit virtual address, ABI-computed slot count $\nslots$, and escaped text (preserving explicit \texttt{\textbackslash n} line boundaries).

\begin{table}[t]
\caption{Argument wire-slot encoding contract}
\label{tbl:slots}
\centering
\footnotesize
\begin{tabular}{@{}llp{3.5cm}@{}}
\toprule
\textbf{Specifier} & \textbf{Slots} & \textbf{Wire Encoding} \\
\midrule
\texttt{\%d}, \texttt{\%i}            & 1 & Sign-extended to 32 bits \\
\texttt{\%u}, \texttt{\%x}, \texttt{\%o}, \texttt{\%c}    & 1 & Zero-extended to 32 bits \\
64-bit integer (\texttt{\%llx}), ptr  & 2 & Low 32-bit word first, then high word \\
Dynamic string (\texttt{\%s})         & $\lceil (\ell{+}1)/4 \rceil$ & Packed 4 chars/word; \texttt{0xFE} sentinel \\
Literal percent (\texttt{\%\%})       & 0 & Consumes no argument slot \\
\bottomrule
\end{tabular}
\end{table}

\subsection{Zero-simulation-time SystemVerilog host decoder}
\label{sec:decoder}

While emitting the token and $\nslots$ argument words incurs $(1+\nslots)\cdot(\Cwr + \Cbar) + \Cpush(\nargs)$ cycles on the simulated processor, string reconstruction on the host server executes in \emph{zero simulated clock cycles} ($\Delta t_{\mathrm{sim}} = 0$). On every 32-bit MMIO write to the tube aperture, the SystemVerilog monitor executes a three-tier priority state machine:
\begin{enumerate}
  \item \textbf{Priority 1 --- In-flight argument collection ($\nslots > 0$):} Incoming 32-bit words are unconditionally captured as raw argument payload and $\nslots$ is decremented.
  \item \textbf{Priority 2 --- Token header lookup ($\nslots = 0$, dictionary hit):} If idle and the 32-bit word matches a valid virtual address in the aperture's dictionary array, a new token frame begins (emitting immediately in $\Delta t_{\mathrm{sim}} = 0$ if $\nslots = 0$, or collecting $\nslots$ argument words).
  \item \textbf{Priority 3 --- Legacy passthrough fallback ($\nslots = 0$, dictionary miss):} Unmatched idle writes fall through to the legacy character handler, supporting mixed builds.
\end{enumerate}

\noindent\textbf{Payload isolation and mixed-build hazards.}
Giving in-flight argument collection highest priority prevents binary argument values (e.g., an integer equal to \texttt{0x04}, the ASCII end-of-test marker, or another string's VA) from triggering false control actions or address lookups while $\nslots > 0$; dynamic strings enforce a $64$-word ($256$-byte) ceiling. In mixed legacy/tokenized builds, two idle-state ($\nslots=0$) hazards exist: (1) an un-tokenized 32-bit word on the word-packed path could numerically collide with a \texttt{.rodata} VA (bounded by per-aperture dictionary isolation and distinct memory regions: zero-extended bytes \texttt{0x00}--\texttt{0xFF} or packed ASCII \texttt{0x20202020}--\texttt{0x7E7E7E7E} vs.\ \texttt{.rodata} in low ROM or high DRAM \texttt{0x80000000}+), and (2) passing a non-literal runtime buffer \texttt{printf(msg)} sends a stack/heap pointer to Priority 3, where a byte equal to \texttt{0x04} can trigger early end-of-test. Placing format literals in a dedicated non-allocatable \texttt{.log\_strings} ELF section (Section~\ref{sec:discussion}) eliminates both hazards at compile time.

\section{Implementation}
\label{sec:impl}

The host-side tokenization architecture comprises three modular components shared across all SoC processor subsystems:
\begin{enumerate}
  \item \textbf{Post-link ELF dictionary extractor:} A standalone Python build hook (using standard ELF section tables~\cite{elfabi} via \texttt{pyelftools}) parses each compiled firmware binary's \texttt{.rodata} section, extracts string literals with their link-time virtual addresses and ABI slot counts $\nslots$, and emits \DICTFILENAME{} without compiler plugins or linker-script edits.
  \item \textbf{ISA-agnostic firmware C encoder:} A compact C99 variadic routine (\texttt{<stdarg.h>}) containing zero inline assembly walks the format string's specifiers in a single pass ($\Cpush$) and issues $1+\nslots$ 32-bit volatile MMIO stores to the core's simulation tube aperture. Each subsystem provides a one-line header alias guarded by \TOKDEFINE{}.
  \item \textbf{SystemVerilog testbench decoder:} A parameterized IEEE 1800~\cite{ieee1800} SystemVerilog module instantiated per tube aperture (\texttt{TOKENIZED\_MODE}) loads the sidecar dictionary into an associative array at simulation start, captures 32-bit MMIO writes, and emits formatted messages into the UVM log stream in zero simulated clock cycles ($\Delta t_{\mathrm{sim}} = 0$). Because all components use standard C99, ELF~\cite{elfabi}, and behavioral SystemVerilog~\cite{ieee1800}, the design is portable to compiled-code simulators such as Verilator~\cite{verilator} and SystemC~\cite{ieee1666} environments.
\end{enumerate}

\section{Empirical Evaluation}
\label{sec:eval}

\begin{table*}[t]
\centering
\footnotesize
\setlength{\tabcolsep}{3.5pt}
\caption{Production Regression Performance Across \SUITEA{} ($N=169$) and \SUITEB{} ($N=689$) ($858$ Matched Full-Chip Test Pairs, $1{,}716$ Runs)}
\label{tab:suite_summary}
\begin{tabular}{@{}lrrrrrrr@{}}
\toprule
\textbf{Metric / Dimension} & \textbf{Mean $\Delta$} & \textbf{Base Med [IQR]} & \textbf{Tok Med [IQR]} & \textbf{Med $\Delta$} & \textbf{GMean Ratio} & \textbf{Best Test} & \textbf{Suite Total} \\
\midrule
\multicolumn{8}{@{}l}{\emph{Panel A: \SUITEA{} --- High-Bandwidth I/O \& Coherency Subsystem ($N=169$ matched pairs, $338$ runs)}} \\
\midrule
Simulated Time (ms)                 & -30.08\% & 2.876 [1.842] & 1.912 [1.255] & -33.32\% & 0.690$\times$ & -78.44\% (4.64$\times$) & 180.52 ms (-31.51\%) \\
Simulator CPU Time (h)              & -36.72\% & 5.56 [6.74]   & 3.09 [4.69]   & -41.51\% & 0.586$\times$ & -83.15\% (5.94$\times$) & 564.53 h (-40.23\%) \\
Wall-Clock Time (h)$^{\dagger}$     & -29.93\% & 6.76 [7.34]   & 4.11 [5.51]   & -32.89\% & 0.662$\times$ & -78.33\% (4.62$\times$) & 557.10 h (-34.50\%) \\
Total Run Dir Disk (MB)$^{\ddagger}$ & -23.65\% & 439.25 [303.6] & 307.57 [228.5] & -25.63\% & 0.757$\times$ & -70.39\% (3.38$\times$) & 22.24 GB (-23.95\%) \\
\quad Instruction Trace (MB)        & -41.38\% & 81.04 [70.64] & 41.05 [49.52] & -42.84\% & 0.565$\times$ & -79.78\% (4.94$\times$) & 6.91 GB (-41.76\%) \\
\quad Bus Transaction Log (MB)      & -36.98\% & 32.99 [27.65] & 18.77 [21.00] & -38.17\% & 0.613$\times$ & -83.85\% (6.19$\times$) & 2.51 GB (-37.55\%) \\
\quad Waveform Dump (MB)            & -28.10\% & 211.45 [99.64] & 136.57 [85.13] & -28.82\% & 0.707$\times$ & -77.24\% (4.39$\times$) & 11.57 GB (-29.84\%) \\
\quad All Log Files (MB)            & -18.16\% & 128.46 [159.7] & 104.93 [113.5] & -18.93\% & 0.812$\times$ & -59.86\% (2.49$\times$) & 6.55 GB (-17.31\%) \\
Norm.\ Throughput (per wall-s)      & +8.51\%  & 1.000$\times$ & 1.060$\times$ & +2.61\%  & 1.043$\times$ & +113.87\% (2.14$\times$) & --- \\
Norm.\ Throughput (per CPU-s)       & +24.02\% & 1.000$\times$ & 1.166$\times$ & +15.05\% & 1.177$\times$ & +147.63\% (2.48$\times$) & --- \\
Norm.\ Simulator RAM                & 0.00\%   & 1.000$\times$ & 1.000$\times$ & 0.00\%   & 1.000$\times$ & -0.01\% (1.00$\times$)   & exact parity \\
\midrule
\multicolumn{8}{@{}l}{\emph{Panel B: \SUITEB{} --- Multi-Core Control \& Boot Subsystem ($N=689$ matched pairs, $1{,}378$ runs)}} \\
\midrule
Simulated Time (ms)                 & -11.38\% & 2.130 [1.730] & 1.793 [1.233] & -12.43\% & 0.880$\times$ & -54.35\% (2.19$\times$) & 260.73 ms (-14.10\%) \\
Simulator CPU Time (h)              & -14.41\% & 3.05 [3.93]   & 2.47 [2.75]   & -21.33\% & 0.790$\times$ & -74.69\% (3.95$\times$) & 750.32 h (-20.73\%) \\
Wall-Clock Time (h)$^{\dagger}$     & -16.74\% & 4.28 [4.85]   & 3.18 [2.92]   & -22.10\% & 0.768$\times$ & -75.77\% (4.13$\times$) & 1{,}113.65 h (-24.47\%) \\
Post-Cleanup Dir Disk (MB)          & -1.49\%  & 41.34 [7.20]  & 41.74 [8.18]  & -0.29\%  & 0.983$\times$ & -28.67\% (1.40$\times$) & 1.90 GB (-0.91\%) \\
Norm.\ Throughput (per wall-s)      & +23.25\% & 1.000$\times$ & 1.124$\times$ & +12.41\% & 1.145$\times$ & +262.21\% (3.62$\times$) & --- \\
Norm.\ Throughput (per CPU-s)       & +19.60\% & 1.000$\times$ & 1.109$\times$ & +10.85\% & 1.113$\times$ & +211.96\% (3.12$\times$) & --- \\
Norm.\ Simulator RAM                & -1.02\%  & 1.000$\times$ & 0.985$\times$ & -1.47\%  & 0.990$\times$ & -1.48\% (0.99$\times$)   & -1.02\% mean \\
\bottomrule
\end{tabular}

\vspace{2pt}
\parbox{\textwidth}{\footnotesize $^{\dagger}$Wall-clock times are measured on a shared production compute cluster and are therefore subject to contention from unrelated jobs; they are reported for operational context and should not be treated as controlled measurements. $^{\ddagger}$The indented artifact classes are \emph{not} disjoint partitions of the run directory and do not sum to the run-directory total; they are reported individually because they are the classes verification engineers manage separately.}
\end{table*}

\subsection{Experimental methodology and statistical rigor}

While the single-call logging reduction ($63.5\%\text{--}90.9\%$ of on-core cycles, Table~\ref{tbl:trace_decomp}) and critical-path savings ($\Delta C_{\mathrm{sim}} = 15.06\,\mu\text{s}$ print-weighted, $15.21\,\mu\text{s}$ unweighted) can be reproduced on a single standalone testbench run, we evaluate host-side tokenized logging at scale across \textbf{$858$ matched full-chip RTL simulation test pairs ($1{,}716$ paired runs, comprising all paired tests that completed cleanly in both legs across $1{,}972$ total regression runs)} spanning two production verification suites on \CHIP{}:
\begin{enumerate}
  \item \textbf{\SUITEA{} --- High-Bandwidth I/O \& Coherency Subsystem ($N=169$ matched pairs, $338$ runs):} A paired A/B production regression of $N=169$ matched full-chip address translation, cache coherency, interrupt delivery, and power-aware (UPF) verification tests. Each test executes twice under identical conditions (identical RTL database, simulator binary [\SIMULATOR{}], random seed, and compute host class), differing solely in whether firmware tokenization (\TOKDEFINE{}) and the SystemVerilog host decoder are enabled.
  \item \textbf{\SUITEB{} --- Multi-Core Control \& Boot Subsystem ($N=689$ matched pairs, $1{,}378$ runs):} A paired A/B production regression of $N=689$ matched full-chip multi-core boot, inter-processor communication (IPC), peripheral serial bus, and power-management tests exercising \CORELIST{}.
\end{enumerate}

Across the $858$ matched pairs, baseline simulation consumed $5{,}022.35$ simulator CPU hours and $6{,}166.43$ wall-clock hours, whereas tokenized simulation consumed $3{,}707.50$ CPU hours and $4{,}495.68$ wall-clock hours---saving $1{,}314.85$ CPU hours ($-26.18\%$), $1{,}670.75$ wall-clock hours ($-27.09\%$), and $24.14\,\text{GB}$ of cluster disk storage in a single regression pass. Because the suites execute in parallel across cluster hosts, these aggregate hours measure datacenter compute resource consumption rather than single-engineer wait time. To characterize right-skewed regression distributions ($\sim 1\text{--}8\,\text{ms}$ simulated time), we report: (1) \textbf{Arithmetic Mean ($\mu$) and Total Sum ($\Sigma = N \cdot \mu$)} for aggregate datacenter compute and storage savings; (2) \textbf{Median (P50) and Interquartile Range (IQR)} for typical single-test turnaround; and (3) \textbf{Geometric Mean (GMean)} for normalized ratios ($\text{Tokenized}/\text{Baseline}$ and speedup $\times$).

\subsection{\texorpdfstring{100\%}{100\%} bit-exact output parity and simulator memory footprint}

Across all $858$ matched full-chip test pairs ($169/169$ in \SUITEA{} and $689/689$ in \SUITEB{}), host-side tokenized logging achieved $100\%$ character-for-character, line-for-line log parity against the unmodified baseline character streams, with all UVM scoreboards, bus monitors, and end-of-test adjudicators passing unmodified. To verify requirement \textbf{R3} (strictly opt-in build parity), we compared compiled firmware artifacts across subsystems built with tokenization disabled: loadable binary images (\texttt{test.bin}, $11{,}040$ and $5{,}536$ bytes), disassembly listings (\texttt{test.dis}), linker maps (\texttt{test.map}), and all $13$ loadable code, data, and DWARF frame sections in \texttt{test.elf} were \textbf{$100\%$ byte-for-byte SHA-256 identical} between baseline and tokenized builds (differing only in the working-directory path string inside non-loadable \texttt{.debug\_str}).

Monitoring simulator process memory (elaborated data-structure and runtime heap footprint reported by \SIMULATOR{} at end of run) confirmed zero memory overhead from the SystemVerilog associative dictionary table. Across \SUITEA{} ($N=169$), where baseline and tokenized runs executed the exact same compiled simulator executable, normalized memory ($\text{Tokenized}/\text{Baseline}$) was $1.0000\times$ across mean, median, and geometric mean ($-0.002\%$ median delta, largest single-test variation $-0.01\%$). Across \SUITEB{} ($N=689$), normalized memory was $0.9898\times$ mean ($-1.02\%$) and $0.9853\times$ median ($-1.47\%$); trace audit confirmed this small reduction reflects static elaboration differences between the baseline snapshot and tokenized workspace build ($-1.47\%$ on the $58.3\%$ standard-RTL subset and $-0.39\%$ on the $41.7\%$ post-DFT/UPF subset, yielding the exact $-1.02\%$ weighted mean).

\subsection{Production regression suite performance (\texorpdfstring{$N=169$}{N=169} and \texorpdfstring{$N=689$}{N=689})}

Table~\ref{tab:suite_summary} summarizes performance across all $N=169$ matched pairs in \SUITEA{} and $N=689$ matched pairs in \SUITEB{}. Non-parametric two-sided Wilcoxon signed-rank tests and $10{,}000$-resample $95\%$ bootstrap confidence intervals confirm high statistical significance across every primary metric:
\begin{itemize}
  \item \textbf{\SUITEA{} ($N=169$):} Median paired reductions are \textbf{$-41.51\%$ in CPU time} ($95\%$ CI $[-46.42\%, -35.69\%]$, $z = -10.43$, $p = 1.73 \times 10^{-25}$), \textbf{$-32.89\%$ in wall time} ($95\%$ CI $[-38.82\%, -29.29\%]$, $z = -10.14$, $p = 3.70 \times 10^{-24}$), \textbf{$-33.32\%$ in simulated time} ($95\%$ CI $[-35.01\%, -31.58\%]$, $W_- = 0$ across all $168$ non-zero pairs, $z = -11.24$, $p = 2.56 \times 10^{-29}$), and \textbf{$-25.63\%$ in run-directory disk} ($95\%$ CI $[-27.23\%, -23.76\%]$, $z = -11.27$, $p = 1.78 \times 10^{-29}$).
  \item \textbf{\SUITEB{} ($N=689$):} Median paired reductions are \textbf{$-21.33\%$ in CPU time} ($95\%$ CI $[-24.30\%, -18.69\%]$, $z = -14.32$, $p = 1.66 \times 10^{-46}$), \textbf{$-22.10\%$ in wall time} ($95\%$ CI $[-26.80\%, -21.11\%]$, $z = -15.95$, $p = 2.83 \times 10^{-57}$), and \textbf{$-12.43\%$ in simulated time} ($95\%$ CI $[-13.16\%, -11.08\%]$, $z = -18.08$, $p = 4.41 \times 10^{-73}$).
\end{itemize}

Three key empirical findings emerge across the two suites:

\paragraph{Aggregate cluster compute and turnaround-time savings}
On \SUITEA{} ($N=169$), simulator CPU consumption drops from $1{,}403.31$ to $838.78\,\text{CPU hours}$ (\textbf{$-564.53\,\text{CPU hours}$, $-40.23\%$}) and wall-clock time drops by \textbf{$-557.10\,\text{hours}$ ($-34.50\%$)}; median single-test CPU time falls from $5.56$ to $3.09\,\text{hours}$ (\textbf{$-41.51\%$}, $1.706\times$ GMean speedup) and wall time falls from $6.76$ to $4.11\,\text{hours}$ (\textbf{$-32.89\%$}, $-2.65\,\text{hours}$ per test). On \SUITEB{} ($N=689$), simulator CPU consumption drops from $3{,}619.04$ to $2{,}868.72\,\text{CPU hours}$ (\textbf{$-750.32\,\text{CPU hours}$, $-20.73\%$}) and wall-clock time drops by \textbf{$-1{,}113.65\,\text{hours}$ ($-24.47\%$)}, with median CPU and wall times falling by \textbf{$-21.33\%$} and \textbf{$-22.10\%$}. Combined ($N=858$), a single pass saves \textbf{$1{,}314.85\,\text{CPU hours}$} and \textbf{$1{,}670.75\,\text{wall-clock hours}$}.

\paragraph{Host CPU time falls faster than simulated time}
Simulated clock cycles drop by $-33.32\%$ (\SUITEA{}) and $-12.43\%$ (\SUITEB{}) at the median, whereas host CPU time drops by $-41.51\%$ and $-21.33\%$ ($+8.19$ and $+8.90$ percentage-point gaps). Normalized simulator throughput ($R_{\mathrm{eff,tok}}/R_{\mathrm{eff,base}}$) rises correspondingly: on \SUITEA{}, $+24.02\%$ mean per CPU-s ($1.177\times$ GMean, $+15.05\%$ median, up to $+147.63\%$) and $+8.51\%$ mean per wall-s ($1.043\times$ GMean, $+2.61\%$ median, up to $+113.87\%$, tempered by the $\approx 1.25\,\text{hour}$ fixed job setup and waveform/trace flush overhead per run); on \SUITEB{}, $+19.60\%$ mean per CPU-s ($1.113\times$ GMean, $+10.85\%$ median, up to $+211.96\%$) and $+23.25\%$ mean per wall-s ($1.145\times$ GMean, $+12.41\%$ median, up to $+262.21\%$). As established in Section~\ref{sec:costmodel}, this reflects the removal of high-fanout fabric and trace-I/O cycles rather than an independent super-linear kernel speedup.

\paragraph{In-production ablation of formatting and per-character barriers versus word-packed transport}
The difference between the two suites reflects their distinct legacy baselines (Equation~\eqref{eq:clegacy}). \SUITEA{}'s primary controller formats messages on-core via \texttt{vsprintf} ($\Cfmt > 0$) and issues one character per bus write with a per-character data-synchronization barrier ($N_{\mathrm{wr}} = N_{\mathrm{bar}} = \msglen$), while application cores format on-core and issue byte-serial cross-fabric stores; tokenization eliminates on-core formatting, collapses $\msglen \rightarrow (1+\nslots)$ writes, and replaces $\msglen$ barriers with one trailing barrier. \SUITEB{}'s primary controller performs no on-core integer formatting ($\Cfmt = 0$)---packing four format-string characters per $32$-bit write alongside framing and raw argument words ($N_{\mathrm{wr}} = \lceil \msglen/4 \rceil + \nargs + 4$, $N_{\mathrm{bar}} = 0$, $\approx 3.4\,\mu\text{s}$ at $1\,\text{GHz}$)---so tokenization eliminates the byte-packing loop and reduces bus writes by $6.7\text{--}7.5\times$ ($15 \rightarrow 2$ and $20 \rightarrow 3$) rather than $15\text{--}18.3\times$ ($30 \rightarrow 2$ and $55 \rightarrow 3$). This predicts the exact ordering observed across the $858$ pairs ($-41.51\%$ vs.\ $-21.33\%$ median CPU time; $-33.32\%$ vs.\ $-12.43\%$ median simulated cycles), showing that even a subsystem without on-core integer formatting gains over $20\%$ in median CPU and wall time purely from removing byte packing and collapsing MMIO writes.

\subsection{Deep-dive production case studies}

To illustrate performance across varying print densities, processor mixes, and simulation modes (standard RTL, low-power UPF, and full waveform dumping), Table~\ref{tab:case_studies} details nine representative production workloads spanning both suites (\CASEA{} through \CASEF{} in \SUITEA{}, multi-core boot/crypto workloads \CASEG{}--\CASEH{} in \SUITEB{}, and peripheral-controller workload \CASEI{} in \SUITEB{}).

\begin{table*}[t]
\centering
\footnotesize
\setlength{\tabcolsep}{3pt}
\caption{Deep-Dive Production Case Studies Across \SUITEA{} (Workloads A1--A6) and \SUITEB{} (Workloads B1--B3)}
\label{tab:case_studies}
\begin{tabular}{@{}lrrrrrrrrrr@{}}
\toprule
 & & \multicolumn{3}{c}{\textbf{Simulated Device Time}} & \multicolumn{3}{c}{\textbf{Simulator CPU Time}} & \textbf{Wall} & \multicolumn{2}{c}{\textbf{Derived per print}$^{*}$} \\
\cmidrule(lr){3-5} \cmidrule(lr){6-8} \cmidrule(lr){9-9} \cmidrule(l){10-11}
\textbf{Case Study / Workload} & \textbf{Prints} & \textbf{Base} & \textbf{Token} & \textbf{$\Delta$} & \textbf{Base (h)} & \textbf{Token (h)} & \textbf{$\Delta$} & \textbf{$\Delta$} & \textbf{$\Delta C_{\mathrm{sim}}$} & \textbf{$\Delta T_{\mathrm{CPU}}$} \\
\midrule
\CASEA{} & 433 & 8.26 ms & 1.78 ms & -78.4\% & 15.83 & 2.67 & -83.2\% & -78.3\% & 14.96 $\mu$s & 109.4 s \\
\CASEB{} & 106 & 3.19 ms & 1.64 ms & -48.6\% & 7.24  & 1.90 & -73.8\% & -65.6\% & 14.64 $\mu$s & 181.6 s \\
\CASEC{} & 81  & 3.23 ms & 1.73 ms & -46.3\% & 7.57  & 2.04 & -73.1\% & -63.6\% & 18.44 $\mu$s & 246.1 s \\
\CASED{} & 85  & 2.52 ms & 1.24 ms & -50.9\% & 4.14  & 1.24 & -70.0\% & -54.6\% & 15.06 $\mu$s & 122.9 s \\
\CASEE{} & 96  & 4.05 ms & 2.79 ms & -31.2\% & 18.26 & 6.49 & -64.5\% & -61.6\% & 13.17 $\mu$s & 441.5 s \\
\CASEF{} & 49  & 2.38 ms & 1.63 ms & -31.6\% & 4.34  & 2.71 & -37.5\% & -34.5\% & 15.35 $\mu$s & 119.5 s \\
\CASEG{} & 312 & 6.84 ms & 2.15 ms & -68.6\% & 11.25 & 2.85 & -74.7\% & -73.2\% & 15.03 $\mu$s & 96.9 s \\
\CASEH{} & 148 & 4.19 ms & 1.97 ms & -53.0\% & 2.83  & 1.27 & -55.0\% & -75.8\% & 15.00 $\mu$s & 37.8 s \\
\CASEI{} & 84  & 2.06 ms & 1.62 ms & -21.0\% & 3.04  & 1.23 & -59.5\% & -53.4\% & 5.14 $\mu$s  & 77.4 s \\
\bottomrule
\end{tabular}

\vspace{2pt}
\parbox{\textwidth}{\footnotesize $^{*}$Derived full-chip test ratios: $\Delta C_{\mathrm{sim}} = (T_{\mathrm{sim,base}} - T_{\mathrm{sim,tok}})/\text{Prints}$ and $\Delta T_{\mathrm{CPU}} = (T_{\mathrm{CPU,base}} - T_{\mathrm{CPU,tok}})/\text{Prints}$. In multi-core boot/concurrency workloads (\CASEA{}--\CASEH{}), critical-path completion is gated by cross-fabric application-core prints ($\Delta C_{\mathrm{sim}} \approx 13.2\text{--}18.4\,\mu\text{s}$/print). In peripheral-dominated \SUITEB{} workloads such as \CASEI{} ($84$ total prints: $58$ word-packed control-subsystem prints [$69\%$] plus $26$ application-core boot prints; adjacent UART/IPC tests span $3.64\text{--}5.41\,\mu\text{s}$/print), $\Delta C_{\mathrm{sim}} = 5.14\,\mu\text{s}$/print is consistent with its $69\%$ word-packed print mix ($5.66\text{--}6.33\,\mu\text{s}$ unweighted single-call mean, reduced to $5.14\,\mu\text{s}$ [an effective $1.23\,\mu\text{s}$/print word-packed contribution] by partial overlap of concurrent control-core prints with off-core bus activity), explaining \SUITEB{}'s $-12.43\%$ median simulated-time and $-21.33\%$ median CPU-time reduction.}
\end{table*}

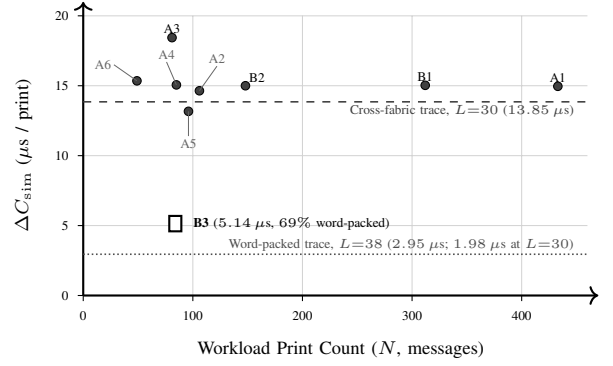
\begin{figure}[!t]
\centering
\begin{tikzpicture}[font=\scriptsize, x=0.145cm, y=0.185cm]
  \draw[gray!35, very thin] (0,0) grid[xstep=10, ystep=5] (46,20);
  \draw[->, thick] (0,0) -- (47,0);
  \draw[->, thick] (0,0) -- (0,21);
  \foreach \x/\lbl in {0/0, 10/100, 20/200, 30/300, 40/400} {
    \draw (\x,0) -- (\x,-0.4) node[below, font=\tiny] {\lbl};
  }
  \foreach \y in {0, 5, 10, 15, 20} {
    \draw (0,\y) -- (-0.5,\y) node[left, font=\tiny] {\y};
  }
  \node[below, font=\scriptsize] at (23.5,-2.6) {Workload Print Count ($N$, messages)};
  \node[rotate=90, above, font=\scriptsize] at (-3.8,10) {$\Delta C_{\mathrm{sim}}$ ($\mu\text{s}$ / print)};

  \draw[dashed, black!75, semithick] (0,13.85) -- (46,13.85)
    node[pos=0.98, below left, font=\tiny, inner sep=1pt] {Cross-fabric trace, $\msglen{=}30$ ($13.85\,\mu\text{s}$)};
  \draw[densely dotted, black!70, semithick] (0,2.95) -- (46,2.95)
    node[pos=0.98, above left, font=\tiny, inner sep=1pt] {Word-packed trace, $\msglen{=}38$ ($2.95\,\mu\text{s}$; $1.98\,\mu\text{s}$ at $\msglen{=}30$)};

  \foreach \px/\py in {
    43.3/14.96,
    10.6/14.64,
    8.1/18.44,
    8.5/15.06,
    9.6/13.17,
    4.9/15.35,
    31.2/15.03,
    14.8/15.00%
  } {
    \filldraw[fill=black!75, draw=black] (\px,\py) circle (1.6pt);
  }
  \node[above, font=\tiny, inner sep=1.5pt] at (43.3,14.96) {A1};
  \node[above, font=\tiny, inner sep=1.5pt] at (31.2,15.03) {B1};
  \node[above right, font=\tiny, inner sep=1.2pt] at (14.8,15.00) {B2};
  \node[above, font=\tiny, inner sep=1.5pt] at (8.1,18.44) {A3};
  \draw[black!60, very thin] (4.9,15.35) -- (2.8,16.5) node[left, font=\tiny, inner sep=1pt] {A6};
  \draw[black!60, very thin] (8.5,15.06) -- (7.6,16.7) node[above, font=\tiny, inner sep=1pt] {A4};
  \draw[black!60, very thin] (10.6,14.64) -- (12.2,16.4) node[above, font=\tiny, inner sep=1pt] {A2};
  \draw[black!60, very thin] (9.6,13.17) -- (9.6,11.4) node[below, font=\tiny, inner sep=1pt] {A5};

  \filldraw[fill=white, draw=black, thick] (8.4-0.55,5.14-0.55) rectangle (8.4+0.55,5.14+0.55);
  \node[right, font=\tiny] at (9.2,5.14) {\textbf{B3} ($5.14\,\mu\text{s}$, $69\%$ word-packed)};
\end{tikzpicture}
\vspace{-5pt}
\caption{Critical-path simulated time saved per print ($\Delta C_{\mathrm{sim}}$) versus print count plotted at exact Table~\ref{tab:case_studies} coordinates across all nine instrumented full-chip workloads, compared against the single-call trace baselines from Table~\ref{tbl:trace_decomp}. Multi-core workloads (\CASEA{}--\CASEH{}, filled circles) cluster around the cross-fabric trace cost ($13.17\text{--}18.44\,\mu\text{s}$/print, unweighted mean $15.21\,\mu\text{s}$, print-weighted mean $15.06\,\mu\text{s}$), whereas peripheral-dominated \CASEI{} (open square, $5.14\,\mu\text{s}$/print) is consistent with its $69\%/31\%$ mix of word-packed ($\approx 2.95\,\mu\text{s}$) and cross-fabric boot prints ($\approx 13.85\,\mu\text{s}$), whose unweighted blended mean is $5.66\text{--}6.33\,\mu\text{s}$, reduced to $5.14\,\mu\text{s}$ by concurrent bus overlap.}
\label{fig:case_scatter}
\end{figure}

Three architectural insights stand out from Table~\ref{tab:case_studies} and Figure~\ref{fig:case_scatter}:
\begin{itemize}
  \item \emph{Up to $5.94\times$ single-test speedup on diagnostic-dense tests.} On \CASEA{}, tokenization reduces simulated device time by $-78.4\%$ ($4.64\times$) and simulator CPU time from $15.83$ to $2.67$ hours ($-83.2\%$, $5.94\times$, saving $13.16$ CPU hours on one test); wall-clock runtime falls from $17.34$ to $3.76$ hours. On \CASEG{} in \SUITEB{}, CPU time drops from $11.25$ to $2.85$ hours ($-74.7\%$, $3.95\times$).
  \item \emph{Direct reconciliation of multi-core ($\approx 15.1\text{--}15.2\,\mu\text{s}$/print) and peripheral-dominated ($\approx 3.6\text{--}5.4\,\mu\text{s}$/print) savings.} Across \CASEA{}--\CASEH{} (Figure~\ref{fig:case_scatter}), where cross-fabric application-core prints sit on the critical path, $\Delta C_{\mathrm{sim}}$ clusters tightly between $13.17\,\mu\text{s}$ and $18.44\,\mu\text{s}$ ($15.21\,\mu\text{s}$ unweighted, $15.06\,\mu\text{s}$ print-weighted; $15.27\,\mu\text{s}$ / $15.08\,\mu\text{s}$ across \SUITEA{} \CASEA{}--\CASEF{}, within $\sim 10\%$ of Table~\ref{tbl:trace_decomp}'s $13.85\,\mu\text{s}$ trace measurement). Conversely, on \CASEI{} ($69\%$ word-packed control-core prints), $\Delta C_{\mathrm{sim}}$ is $5.14\,\mu\text{s}$/print ($-21.0\%$ simulated time, $-59.5\%$ CPU time), bridging the single-call word-packed model ($\approx 2.95\,\mu\text{s}$) and the suite-wide \SUITEB{} median ($-12.43\%$ simulated time, $-21.33\%$ CPU time).
  \item \emph{Large CPU savings in power-aware (UPF) simulation.} On \CASEE{}, run in Unified Power Format low-power mode where power-domain instrumentation raises per-cycle evaluation cost, the $13{,}170$ logging cycles attributable to a single \texttt{printf} consume $441.5$ server CPU seconds ($7.36$ CPU minutes). Tokenization saves $11.77$ CPU hours ($-64.5\%$) on this single test.
\end{itemize}

\subsection{Verification artifact and full-chip waveform compression}

Because full-chip regressions execute hundreds of tests nightly, simulation disk footprint is a critical datacenter bottleneck. Table~\ref{tab:storage} summarizes artifact compression across full regression runs ($N=169$ and $N=689$) and deep debug waveform dump runs.

\begin{table}[t]
\centering
\footnotesize
\setlength{\tabcolsep}{4pt}
\caption{Verification Artifact \& Full-Chip Waveform Disk Compression (\SUITEA{}: $N{=}169$; both suites: $N{=}858$)}
\label{tab:storage}
\begin{tabular}{@{}llrr@{}}
\toprule
\textbf{Scope} & \textbf{Artifact Class} & \textbf{Base $\to$ Tok} & \textbf{Saved} \\
\midrule
Suite A median   & Instruction Trace   & 81.0 $\to$ 41.1 MB    & -42.8\% \\
Suite A median   & Bus Tracker Log     & 33.0 $\to$ 18.8 MB    & -38.2\% \\
Suite A median   & Regression Waveform & 211.5 $\to$ 136.6 MB  & -28.8\% \\
Suite A total    & Suite Run Dirs      & 92.9 $\to$ 70.6 GB    & -24.0\% \\
Both suites      & Total Cluster Disk  & 301.0 $\to$ 276.9 GB  & -8.0\% \\
\midrule
Workload A1      & Instruction Trace   & 264.2 $\to$ 53.4 MB   & -79.8\% \\
Workload A1      & Full Run Directory  & 1015.5 $\to$ 300.7 MB & -70.4\% \\
\midrule
Workload A6 deep & Full-Chip Waveform  & 15.85 $\to$ 9.90 GB   & -37.5\% \\
Boot, deep       & Full-Chip Waveform  & 13.91 $\to$ 10.93 GB  & -21.5\% \\
\bottomrule
\end{tabular}
\end{table}

As shown in Table~\ref{tab:storage}, instruction retirement traces shrink by \textbf{$-42.84\%$ median} across \SUITEA{} ($N=169$, $1.770\times$ GMean, saving $6.91\,\text{GB}$ across the suite and up to \textbf{$-79.78\%$ / $4.94\times$} on \CASEA{}), reflecting the elimination of retired \texttt{libc} formatting instructions. Regression waveform dumps shrink by \textbf{$-28.82\%$ median} ($11.57\,\text{GB}$ saved across \SUITEA{}), while deep full-chip debug waveforms shrink by \textbf{$5.94\,\text{GB}$ ($-37.50\%$)} on \CASEF{} ($15.85 \to 9.90\,\text{GB}$) and \textbf{$2.98\,\text{GB}$ ($-21.45\%$)} on Multi-Core Boot ($13.91 \to 10.93\,\text{GB}$).

Comparing disk behavior across the two suites explains why \SUITEA{} saves $22.24\,\text{GB}$ ($-23.95\%$) whereas \SUITEB{} post-cleanup directories change by $1.90\,\text{GB}$ ($-0.91\%$, $208.1 \to 206.2\,\text{GB}$, median $41.34\,\text{MB}$): \SUITEA{} retains the binary trace subdirectory (waveforms, instruction traces, and bus trackers) for all $169$ passing tests, whereas \SUITEB{} post-pass cleanup deletes the binary trace subdirectory across all $689$ passing tests, leaving only the simulation console log plus run metadata ($610$ standard tests [$88.5\%$, $<100\,\text{MB}$] average $41.29\,\text{MB}$ [$24.60\,\text{GB}$], $7$ intermediate tests [$1.0\%$, $100\text{--}200\,\text{MB}$] total $1.64\,\text{GB}$ [$0.8\%$], and $72$ uncore AXI burst stress tests [$10.4\%$, $>200\,\text{MB}$] dump $8.5\text{--}19.0\,\text{GB}$ of UVM AXI transaction logs into the console log, totaling $179.93\,\text{GB}$ [$87.3\%$]). Because host-side tokenization preserves human-readable text in the simulation console log (\textbf{R1}), post-cleanup text log volume is unchanged while cycle-proportional binary traces shrink by $28.8\%\text{--}42.8\%$.

\section{Architectural Trade-Offs and Discussion}
\label{sec:discussion}

\subsection{Latent timing dependencies exposed by accelerating \texttt{printf}}
\label{sec:latent}

A legacy $\msglen$-character message occupies the simulation tube aperture for $N_{\mathrm{wr}}(\msglen)$ ordered bus transactions ($3.04\text{--}15.23\,\mu\text{s}$ in Table~\ref{tbl:trace_decomp}). When firmware or testbench code is developed in the presence of that delay, unsynchronized multi-core sequences can inadvertently rely on \texttt{printf} as an implicit timing barrier---a latent defect that remains hidden until logging is accelerated or the firmware executes on emulation or silicon.

Hardware trace inspection during deployment illustrates how narrow such race margins can be. In one multi-core test, six processors perform an unsynchronized read-modify-write on a shared configuration register with a diagnostic print inside the read-to-write window. Under the baseline word-packed driver, the writing processor's $38$-character, $1$-argument print issued $15$ ordered 32-bit MMIO writes ($\lceil 38/4 \rceil + 1 + 4 = 15$ at $\approx 227.1\,\text{ns/tx}$), inserting $3{,}407\,\text{ns}$ of delay into an $\approx 8.6\,\mu\text{s}$ race window; a second processor sampled the register inside that window and read the stale value by a margin of just $261\,\text{ns}$ ($3\%$). Under tokenization ($2$ writes, $\approx 454\,\text{ns}$), the register write lands $2.95\,\mu\text{s}$ earlier and immediately exposes the missing lock. Surfacing latent multi-core races before silicon bring-up is a direct verification benefit of removing artificial logging delays.

\subsection{Position-independent code and dedicated \texttt{.log\_strings} sections}

For position-independent code (PIC) or dynamically relocated modules, the encoder can emit $\text{Token} = \text{VA}_{\text{fmt}} - \text{Base}_{\text{module}}$ without call-site edits. While our default extractor scans \texttt{.rodata} so linker scripts require zero edits, placing format literals into a non-allocatable (\texttt{NOLOAD}/\texttt{INFO}) \texttt{.log\_strings} section inside the subsystem macro alias provides compile-time safety:
\begin{center}\scriptsize
\begin{tabular}{@{}l@{}}
\texttt{\#define SUBSYS\_PRINTF(fmt, ...)\ (\{\ static const char \_f[]\textbackslash} \\
\hspace*{0.8em}\texttt{\_\_attribute\_\_((section(".log\_strings"))) = (fmt);\textbackslash} \\
\hspace*{0.8em}\texttt{tok\_printf(\_f, \#\#\_\_VA\_ARGS\_\_);\ \})}
\end{tabular}
\end{center}
Because C requires a \texttt{static const char[]} initializer to be a compile-time literal, non-literal buffers (\texttt{printf(msg)}) fail at \emph{compile time}, eliminating both mixed-build hazards in Section~\ref{sec:decoder} while stripping format strings from target ROM/SRAM. Similarly, deploying our 32-bit slot protocol over hardware emulator and FPGA trace transactors~\cite{firesim2018} removes formatting loops from emulated CPUs and cuts host-transactor bandwidth by $6.7\times\text{--}18.3\times$.

\subsection{Scope and methodological limitations}
\label{sec:threats}

Three methodological boundaries frame our results: (1) \emph{Paired single-seed execution and cohort representativeness:} each of the $858$ matched test pairs ($169/216$ baseline-passing in \SUITEA{} and $689/732$ in \SUITEB{}, drawn from $1{,}972$ total regression runs) ran once per configuration with an identical random seed ($n=1$ per seed). Distributional comparison confirms the evaluated cohort is conservative: across \SUITEB{}, median baseline simulated time across all $732$ baseline-passing runs is $2.151\,\text{ms}$ vs.\ $2.130\,\text{ms}$ on the $689$ paired runs ($<1\%$ delta), while in \SUITEA{} the $47$ \SUITEA{} tests that pass in baseline but not under tokenization (environment timeouts and race-exposed tests, Section~\ref{sec:latent}) contain a higher share of long-running UPF low-power tests ($53.2\%$ [$25/47$] vs.\ $28.4\%$ [$48/169$]; $11.47\,\text{h}$ vs.\ $8.40\,\text{h}$ mean baseline CPU time), which exhibit the largest per-print CPU savings ($441.5\,\text{CPU-s/print}$, \CASEE{}). While Wilcoxon signed-rank tests ($p \le 3.7 \times 10^{-24}$) and $95\%$ bootstrap confidence intervals rule out random suite-level jitter, shared-cluster wall-clock times remain subject to host contention (e.g., \CASEH{} in Table~\ref{tab:case_studies}); (2) \emph{Trace vs.\ whole-test parameterization:} single-call parameters ($\Cfmt, \Cwr, \Cbar, \Cpush$) are measured directly from processor instruction traces in Table~\ref{tbl:trace_decomp}, whereas whole-test $\Delta C_{\mathrm{sim}}$ values in Table~\ref{tab:case_studies} divide test-level runtime savings by print count across nine instrumented workloads; and (3) \emph{Deployed baselines and generality:} our evaluation compares against the three production logging drivers deployed on \CHIP{} (including \SUITEB{}'s host-formatted word-packed driver) rather than an in-SRAM ring buffer or a logging-disabled build, so absolute speedup magnitudes depend on a given SoC's interconnect latency $\Cwr$ and workload print density $\density$.

\section{Conclusion}
\label{sec:conclusion}

Pre-silicon SoC verification pays a high simulation tax for formatted firmware logging: every $30$-character, $1$-argument \texttt{printf} consumes $3.04\text{--}15.23\,\mu\text{s}$ ($19.60\,\mu\text{s}$ at $34$ characters and $4$ arguments; $15.27\,\mu\text{s}$ unweighted / $15.08\,\mu\text{s}$ print-weighted critical-path mean across \SUITEA{}) in formatting instructions ($61.7\%$ on local tubes), barriers, and MMIO writes ($84.7\%$ on cross-fabric paths), costing $37.8\text{--}441.5$ host CPU seconds per message. Using link-time ELF virtual addresses as tokens, an ISA-independent 32-bit MMIO slot protocol across Arm, RISC-V, and DSP cores, and a zero-simulation-time ($\Delta t_{\mathrm{sim}}=0$) SystemVerilog decoder, our architecture removes on-core radix formatting ($\Cfmt \to 0$) and cuts single-call logging latency to $(1+\nslots)\cdot(\Cwr+\Cbar) + \Cpush(\nargs)$---reducing on-core logging cycles by $63.5\%\text{--}90.9\%$ ($2.74\times\text{--}11.03\times$), MMIO bus-stall cycles by $84.6\%\text{--}93.3\%$, retired instructions by $69.6\%\text{--}70.4\%$, and bus transactions by $85.0\%\text{--}94.5\%$ ($6.7\times\text{--}18.3\times$) with zero call-site edits. Across $858$ matched full-chip test pairs ($1{,}716$ paired runs across $1{,}972$ total runs; $N=169$ in \SUITEA{}, $N=689$ in \SUITEB{}), it achieves \textbf{100\% bit-exact output parity} and \textbf{$0.00\%$ simulator RAM overhead}, saves \textbf{$1{,}314.85$ CPU hours ($-26.18\%$)} and \textbf{$1{,}670.75$ wall-clock hours ($-27.09\%$)} per pass, reduces median single-test CPU time by \textbf{$41.51\%$ ($1.706\times$ GMean, up to $5.94\times$)} on \SUITEA{} and \textbf{$21.33\%$ ($1.266\times$ GMean)} on \SUITEB{}, boosts mean effective simulator throughput by \textbf{$+24.02\%$ / $+19.60\%$ per CPU-s ($+8.51\%$ / $+23.25\%$ per wall-s)}, and shrinks \SUITEA{} compressed waveforms by \textbf{$28.82\%$ median ($2.98\text{--}5.94\,\text{GB}$ saved per deep debug test)}.

\bibliographystyle{IEEEtran}
\bibliography{refs}

\end{document}